\documentclass[prd,aps,showpacs,twocolumn,twoside]{revtex4-1}
\usepackage{amsfonts,amsmath,amssymb,bm,graphicx,subfigure}

\begin{document}

\title{Impossibility of the strong magnetic fields generation in an electron-positron plasma}

\author{Maxim Dvornikov}

\email{maxim.dvornikov@usp.br}

\affiliation{Institute of Physics, University of S\~{a}o Paulo, CP 66318, CEP 05315-970 S\~{a}o Paulo, SP, Brazil;
\\
Pushkov Institute of Terrestrial Magnetism, Ionosphere
and Radiowave Propagation (IZMIRAN), \\
142190 Troitsk, Moscow, Russia;
\\
Research School of Physics and Engineering, Australian National University, 2601 Canberra, ACT, Australia}

\date{\today}

\begin{abstract}
We examine the issue whether a magnetic field can be amplified in a background matter consisting of electrons and positrons self-interacting within the Fermi model.
For this purpose we compute the antisymmetric contribution to the photon polarization tensor in this matter having nonzero temperature and chemical potential. It is shown that this contribution is vanishing in the static limit. Then we study a particular case of a degenerate relativistic electron gas present in a magnetar. We demonstrate that a seed magnetic field is attenuated in this case.
Thus, contrary to the recent claim, we show that there is no magnetic
field instability in such a system, which can lead to the magnetic field growth. Therefore recently proposed mechanism cannot be used for the explanation of strong magnetic
fields of magnetars.
\end{abstract}


\maketitle

Recently the new mechanism of the strong magnetic fields generation
in various astrophysical and cosmological plasmas was proposed in
Ref.~\cite{BoyRucSha12}. It is based on the nonzero Chern-Simons (CS)
parameter in the Standard Model plasma consisting of all kinds of neutrinos,
charged leptons, and quarks. This CS parameter results in
the instability of magnetic fields ($B$-fields), which, in its turn, leads to the
growth of a seed magnetic field. As predicted in Ref.~\cite{BoyRucSha12},
there is a very strong magnetic field amplification in case of the electron-positron
($e^{-}e^{+}$) self-interacting plasma, which can be used to explain
strong magnetic fields of magnetars~\cite{DunTho92}. We show, basing on the explicit
calculation of the CS parameter in this system using the
imaginary time perturbation theory, that this result of Ref.~\cite{BoyRucSha12}
is invalid.

In our work we shall examine the evolution of a $B$-field in an isotropic $e^{-}e^{+}$ plasma where macroscopic fluxes are absent. For this purpose we shall calculate one of the photon form factors contributing to the photon polarization tensor $\Pi_{\mu\nu}(x)$. In an isotropic medium $\Pi_{\mu\nu} = \smallint \mathrm{d}^{4}x e^{\mathrm{i}kx} \Pi_{\mu\nu}(x)$ has the form,
\begin{equation}\label{polariztens}
  \Pi_{\mu\nu} =
  \left(
    g_{\mu\nu} - \frac{k_\mu k_\nu}{k^2}
  \right)
  \Pi_1  +
  \mathrm{i}\varepsilon_{\mu\nu\alpha0} k^\alpha
  \Pi_2 +
  \frac{k_\mu k_\nu}{k^2}
  \Pi_3
\end{equation}
where $\Pi_{1,2,3} = \Pi_{1,2,3}(k)$ are the form factors of a photon, $k^\mu = \left(k^{0},\mathbf{k}\right)$ is the photon momentum, $g_{\mu\nu} = \mathrm{diag}(+1, -1, -1, -1)$ is the metric tensor in Minkowski space, and $\varepsilon_{\mu\nu\alpha\beta}$ is the absolute antisymmetric tensor having $\varepsilon^{0123} = +1$. In Eq.~\eqref{polariztens} we adopt the notation for the photon form factors from Ref.~\cite{BoyRucSha12}. Note that a more general form of $\Pi_{\mu\nu}$, which also includes the contributions of nonzero macroscopic plasma flows, is given in Ref.~\cite{MohNiePal98}.

It should be noted that $\Pi_1$ in Eq.~\eqref{polariztens} has a nonzero value even for purely virtual electrons and positrons. It describes the vacuum polarization in QED. The generation of the plasmon mass in a QED plasma takes place if $\Pi_3 \neq 0$. Thus $\Pi_{1,3} \neq 0$ in case of a parity conserving interaction. We shall study the contribution of a parity violating interaction to $\Pi_{\mu\nu}$. That is why we shall concentrate on the analysis of $\Pi_2$, which is absent in the QED case. We also mention that, if $\Pi_2 \neq 0$, the effective Lagrangian of the electromagnetic
field acquires a CS term, $\mathcal{L}_\mathrm{CS} = \Pi_2 (\mathbf{A}\cdot\mathbf{B})$, where $\mathbf{A}$ is the vector potential and $\mathbf{B} = (\nabla \times \mathbf{A})$ is the magnetic field.


Following Ref.~\cite{BoyRucSha12}, let us examine the generation of a CS term in an $e^{-}e^{+}$ plasma, with particles in this plasma self-interacting in frames of the Fermi model. Denoting the electron-positron field as a bispinor $\psi$, we get the
Lagrangian of this system, which also includes the interaction of $\psi$ with the external
electromagnetic field $A^{\mu} = (A^0,\mathbf{A})$, in the form~\cite{MohPal04},
\begin{align}\label{eq:LFermi}
  \mathcal{L}_{\mathrm{I}} = & -e\bar{\psi}\gamma^{\mu}\psi A_{\mu}
  -
  \frac{G_{\mathrm{F}}}{\sqrt{2}}
  \bar{\psi}\gamma_{\alpha}
  \left(
    g_{V}-g_{A}\gamma^{5}
  \right)
  \psi
  \notag
  \\
  & \times
  \bar{\psi}\gamma^{\alpha}
  \left(
    g_{V}-g_{A}\gamma^{5}
  \right)\psi,
\end{align}
where $\gamma^{\mu}=(\gamma^{0},\bm{\gamma})$ are the Dirac matrices,
$\gamma^{5}=\mathrm{i}\gamma^{0}\gamma^{1}\gamma^{2}\gamma^{3}$,
$e$ is the electron's electric charge, $G_{\mathrm{F}}$ is the Fermi
constant, $g_{V}=-\tfrac{1}{2}+2\sin^{2}\theta_{W}$ and $g_{A}=-\tfrac{1}{2}$
are the vector and axial constants of the Fermi interaction, and $\theta_{W}$
is the Weinberg angle. The electroweak interaction in Eq.~\eqref{eq:LFermi} is parity violating. Thus one expects that $\Pi_2$ in Eq.~\eqref{polariztens} could be nonzero.

\begin{figure}
  \centering
  \subfigure[]
  {\label{a}
  \includegraphics[scale=.9]{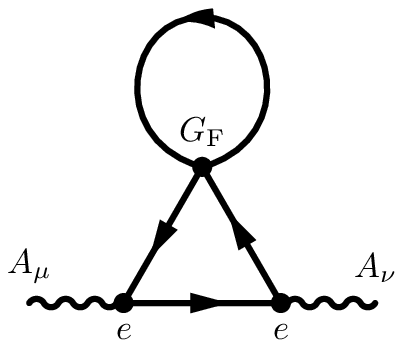}}
  \subfigure[]
  {\label{b}
  \includegraphics[scale=.9]{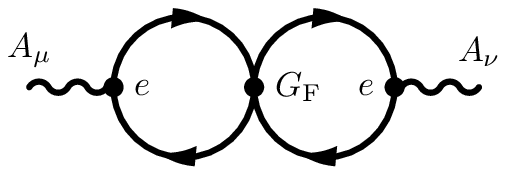}}
  \caption{Schematic representation of the contributions to $\Pi_{\mu\nu}$
  in Eq.~\eqref{eq:Pimunu}. Straight lines stay for $\psi$.
  \label{Feyndiagr}}
\end{figure}

The contributions to $\Pi_{\mu\nu}$, potentially containing antisymmetric terms, are schematically shown in Fig.~\ref{Feyndiagr}.
These terms are quadratic in $e$ and linear in $G_{\mathrm{F}}$. However, since the Lagrangian in Eq.~\eqref{eq:LFermi} includes the self-interaction, one should better rely on the expansion of the $S$-matrix, $S = \mathcal{T} \left[ \exp \left( \mathrm{i} \int\mathrm{d}^{4}x \mathcal{L}_{\mathrm{I}} \right) \right]$,
where the symbol $\mathcal{T}$ stays for the time ordering. Averaging
the aforementioned terms over the state corresponding to the vacuum of $\psi$
and taking into account the definition of
$\Pi_{\mu\nu}(x)$, $\left\langle S\right\rangle_{0} = \frac{\mathrm{i}}{2} \int \mathrm{d}^{4}x \mathrm{d}^{4}y A^{\mu}(x)\Pi_{\mu\nu}(x-y)A^{\nu}(y)$,
we get that
\begin{widetext}
\begin{align}\label{eq:Pimunu}
  \Pi_{\mu\nu} = &
  \sqrt{2}e^{2}G_{\mathrm{F}}
  \int\frac{\mathrm{d}^{4}p\mathrm{d}^{4}q}{(2\pi)^{8}}
  \big\{
  \text{tr}
  \left[
    \gamma_{\mu}S_{\mathrm{F}}(k+p)\gamma_{\nu}
    S_{\mathrm{F}}(p)\Gamma_{\alpha}S_{\mathrm{F}}(p)
  \right]
  \cdot
  \text{tr}
  \left[
    \Gamma^{\alpha}S_{\mathrm{F}}(q)
  \right]
  \nonumber
  \\
  & -
  \text{tr}
  \left[
    \gamma_{\mu}S_{\mathrm{F}}(k+p)\Gamma_{\alpha}
    S_{\mathrm{F}}(k+p)\gamma_{\nu}S_{\mathrm{F}}(p)
  \right]
  \cdot
  \text{tr}
  \left[
    \Gamma^{\alpha}S_{\mathrm{F}}(q)
  \right]
  \nonumber
  \\
  & +
  \text{tr}
  \left[
    \gamma_{\mu}S_{\mathrm{F}}(k+p)\Gamma_{\alpha}S_{\mathrm{F}}(p)
  \right]
  \cdot
  \text{tr}
  \left[
    \gamma_{\nu}S_{\mathrm{F}}(q)\Gamma^{\alpha}S_{\mathrm{F}}(k+q)
  \right]
  \nonumber
  \\
  & +
  \text{tr}
  \left[
    \gamma_{\mu}S_{\mathrm{F}}(k+p)\Gamma_{\alpha}S_{\mathrm{F}}(k+q)
    \gamma_{\nu}S_{\mathrm{F}}(q)\Gamma^{\alpha}S_{\mathrm{F}}(p)
  \right]
  \big\},
\end{align}
\end{widetext}
where $S_{\mathrm{F}}(k) = \smallint\mathrm{d}^{4}xe^{\mathrm{i}kx}S_{\mathrm{F}}(x)=(m- {\not k})^{-1}$
is the Fourier transforms of the vacuum electron
propagator $S_{\mathrm{F}}(x-y)=i\left\langle \mathcal{T}\left[\psi(x)\bar{\psi}(y)\right]\right\rangle_{0}$ and
$m$ is the electron mass. In Eq.~(\ref{eq:Pimunu}) we define $\Gamma_{\alpha}=\gamma_{\alpha}\left(g_{V}-g_{A}\gamma^{5}\right)$
for brevity.


Taking the antisymmetric part of $\Pi_{ij} = \mathrm{i}\varepsilon_{ijn}k^{n}\Pi_{2}$ in Eq.~\eqref{eq:Pimunu} we obtain $\Pi_2$ in Eq.~\eqref{polariztens}. First we calculate $\Pi_{2}$
for purely virtual $\psi$'s. Using the standard methods of QFT we get that $\Pi_{2} = 0$ in this case. Thus there is no ambiguity in the CS term determination, mentioned
in Ref.~\cite{JacKos99}, which can appear in a Fermi-like theory with
a parity violation.

To get the contribution to $\Pi_{2}$ from a $e^{-}e^{+}$ plasma
with the nonzero temperature $T$ and the chemical potential $\mu$ we make the following replacement in Eq.~(\ref{eq:Pimunu})~\cite{KapGal06}:
$\mathrm{i}\int\frac{dp_{0}}{2\pi}\to T\sum_{n}$, where $p_{0}=(2n+1)\pi T\mathrm{i}+\mu$
and $n=0,\pm1,\pm2,\dotsc$. One can show that the contribution to $\Pi_{2}$ arising
from the two last lines in Eq.~(\ref{eq:Pimunu}) cancel each other now. The remaining nonzero contribution can be derived
using the standard technique for the summation over the Matsubara
frequencies (see, e.g., Ref.~\cite{DvoSem14}),
%
\begin{align}\label{eq:P21}
  \Pi_{2}= &
  \frac{
  \left(
    1-4\sin^{2}\theta_{W}
  \right)
  }
  {2\sqrt{2}}
  e^{2}G_{\mathrm{F}}
  \left(
    n_{e}-n_{\bar{e}}
  \right)
  \notag
  \\
  & \times
  \int_{0}^{1}(1-x)\mathrm{d}x
  \int\frac{\mathrm{d}^{3}p}{(2\pi)^{3}}
  \frac{1}{\mathcal{E}_{\mathbf{p}}^{3}}
  \nonumber
  \\
  & \times
  \bigg\{
    \left[
      J'_{1}-J''_{1}
    \right] -
    \left[
      J'_{0}-J''_{0}
    \right]
    \notag
    \\
    & \times
    \frac{3}{\mathcal{E}_{\mathbf{p}}^{2}}
    \left[
      \mathbf{p}^{2}
      \left(
        1-\frac{2}{3}x
      \right) -
      m^{2}(1+x)-k^{2}x^{2}
    \right]
  \bigg\},
\end{align}
where $\mathcal{E}_{\mathbf{p}}=\sqrt{\mathbf{p}^{2}+M^{2}}$, $M^{2}=m^{2}-k^{2}x(1-x)$,
and
\begin{align}\label{eq:J02p}
  J'_{0} = &
  \frac{1}{\exp[\beta(\mathcal{E}_{\mathbf{p}}+\mu')]+1} +
  \frac{\beta\mathcal{E}_{\mathbf{p}}}{2}
    \frac{ 1 + \frac{\beta \mathcal{E}_{\mathbf{p}}}{3}
    \tanh
    \left[
      \frac{\beta}{2}(\mathcal{E}_{\mathbf{p}}+\mu')
    \right]
    }
    {1+\cosh[\beta(\mathcal{E}_{\mathbf{p}}+\mu')]}
    \notag
    \\
    & +
    \left(
      \mu' \to - \mu'
    \right),
  \nonumber
  \\
  J'_{1}= &
  \frac{1}{\exp[\beta(\mathcal{E}_{\mathbf{p}}+\mu')]+1} +
  \frac{\beta\mathcal{E}_{\mathbf{p}}}{2}
    \frac{1-\beta\mathcal{E}_{\mathbf{p}}
    \tanh
    \left[
      \frac{\beta}{2}(\mathcal{E}_{\mathbf{p}}+\mu')
    \right]
    }
    {1+\cosh[\beta(\mathcal{E}_{\mathbf{p}}+\mu')]}
    \notag
    \\
    & +
    \left(
      \mu' \to - \mu'
    \right).
\end{align}
%
Here $\beta=1/T$ and $\mu'=\mu+k_{0}x$. The expressions for $J''_{0,1}$
are analogous to those in Eq.~(\ref{eq:J02p}) if we make the replacement
$\mu'\to\mu''=\mu+k_{0}(1-x)$. The electron and positron number densities
are defined as $n_{e,\bar{e}}=\int\frac{\mathrm{d}^{3}p}{(2\pi)^{3}}
(\exp[\beta(E_{\mathbf{p}}\mp\mu)]+1)^{-1},$
where $E_{\mathbf{p}}=\sqrt{\mathbf{p}^{2}+m^{2}}$ is the electron
energy. To derive Eqs.~(\ref{eq:P21}) and~(\ref{eq:J02p}) we suppose
that $k^{2}<4m^{2}$, i.e. no creation of $e^{-}e^{+}$ pairs occurs~\cite{Bra92}.

It is worth mentioning that the first two lines in Eq.~\eqref{eq:Pimunu} correspond to Fig.~\ref{Feyndiagr}(a) and the last two lines in Eq.~\eqref{eq:Pimunu} -- to Fig.~\ref{Feyndiagr}(b). One can see that the first two lines in Eq.~\eqref{eq:Pimunu} have opposite signs as a consequence of the anticommutativity of $\psi$. This fact is important for the cancelation of divergencies in $\Pi_2$. The direct calculation shows that the contribution to $\Pi_2$ from the last two lines in Eq.~\eqref{eq:Pimunu} equals zero for both virtual and real $\psi$. This fact also results from the possibility to cut the graph in Fig.~\ref{Feyndiagr}(b) by a vertical line passing through the central vertex into two identical parts which are linear in momentum.

Let us express $\Pi_{2}$ in Eq.~(\ref{eq:P21}) as $\Pi_{2}=\tfrac{\alpha_{\mathrm{em}}}{\sqrt{2}\pi}\left(1-4\sin^{2}\theta_{W}\right)G_{\mathrm{F}}\left(n_{e}-n_{\bar{e}}\right)F$,
where $\alpha_{\mathrm{em}}=\tfrac{e^{2}}{4\pi}$ is the fine structure
constant and $F$ is the dimensionless function. We shall analyze
this function in the static limit $k_{0}\to0$. We mention that, if
we neglect $k_{0}$ in Eq.~(\ref{eq:J02p}), $J_{0,1}'=J_{0,1}''$
and $\Pi_{2}\to0$.

To study more carefully the behavior of $\Pi_{2}$ in the static limit we shall consider the case of a degenerate
relativistic electron gas inside a magnetar, where $\mu\gg(m,T)$. The dispersion law for
long waves with $k_{0}\gg|\mathbf{k}|$ in
this background matter reads $k^2 \approx \omega_{p}^{2}$~\cite{DvoSem14},
where $\omega_{p}^{2}=\tfrac{4}{3\pi}\alpha_{\mathrm{em}}\mu^{2}$
and $k_{0}\approx0.06\mu\ll\mu$. Thus $k_{0}/\mu$ is the small parameter and hence can be neglected.

Considering the limit $T/\mu \to 0$ in Eqs.~\eqref{eq:P21} and~\eqref{eq:J02p} and using the results of Ref.~\cite{DvoSem14}, we get $F$ in the explicit form,
\begin{align}\label{Fdeg}
  F = & \int_{0}^{1}(1-x)\mathrm{d}x
  \big[
    Z^{2}(I'_{2} - I''_{2})
    \notag
    \\
    & -
    2(1-x)(I'_{1}-I''_{1}) - (I'_0 - I''_0)
  \big],
\end{align}
where
\begin{widetext}
\begin{align}\label{nachdeg}
  I'_0 = &
  \left\{
    \left[
      2(1-x) - 3\frac{Z^{2}}{\mathcal{M}'^{2}}
    \right]
    \left(
      1-\frac{\tilde{M}^{2}}{\mathcal{M}'^{2}}
    \right)^{1/2}
    +
   \left[
      2\left(1-\frac{x}{3}\right) + \frac{Z^{2}}{\mathcal{M}'^{2}}
      \left(
        1-2\frac{\tilde{M}^{2}}{\mathcal{M}'^{2}}
      \right)
    \right]
    \left(
      1-\frac{\tilde{M}^{2}}{\mathcal{M}'^{2}}
    \right)^{-1/2}
  \right\}
  \notag
  \\
  & \times
  \theta(\mathcal{M}'-\tilde{M})
  \notag
  \\
  I'_{1} = &
  \bigg\{
    \ln
    \left(
      \frac{\mathcal{M}'+\sqrt{\mathcal{M}'^{2}-\tilde{M}^{2}}}{\tilde{M}}
    \right)
	+ \frac{1}{\mathcal{M}'\tilde{M}^{2}}
    \left[
      \left(
        \mathcal{M}'^{2}-\tilde{M}^{2}
      \right)^{3/2} - \mathcal{M}'^{2}\sqrt{\mathcal{M}'^{2}-\tilde{M}^{2}}
    \right]
  \bigg\}
  \theta(\mathcal{M}'-\tilde{M}),
  \notag
  \\
  I'_{2} = &
  \frac{
  \left(
    \mathcal{M}'^{2}-\tilde{M}^{2}
  \right)^{3/2}
  }
  {\mathcal{M}'^{3}\tilde{M}^{2}}\theta(\mathcal{M}'-\tilde{M}),
  \notag
  \\
  \mathcal{M}' =	& 1+\frac{k_{0}}{\mu}x,
  \quad
  \mathcal{M}''=1+\frac{k_{0}}{\mu}(1-x),
  \quad
  \tilde{M} = \frac{1}{\mu}\sqrt{m_{\mathrm{eff}}^{2}-k^{2}x(1-x)},
  \notag
  \\
  Z^{2} = & \tilde{M}^{2}
  \left(
    1-\frac{2}{3}x
  \right) +
  \frac{1}{\mu^{2}}
  \left[
    m_{\mathrm{eff}}^{2}(1+x)+k^{2}x^{2}
  \right],
\end{align}
\end{widetext}
where $\theta(z)$ is the Heaviside step function.
The expressions for $I''_{0,1,2}$ can be obtained by replacing $\mathcal{M}' \to \mathcal{M}''$ in $I'_{0,1,2}$ in Eq.~\eqref{nachdeg}. In Eqs.~\eqref{Fdeg} and~\eqref{nachdeg} we take into account that an electron
acquires the effective mass $m_{\mathrm{eff}}^{2}=\tfrac{e^{2}}{8\pi^{2}}\mu^{2}$ in case of $\mu \gg m$ as found in Ref.~\cite{Bra92}.

\begin{figure}
  \includegraphics[scale=.4]{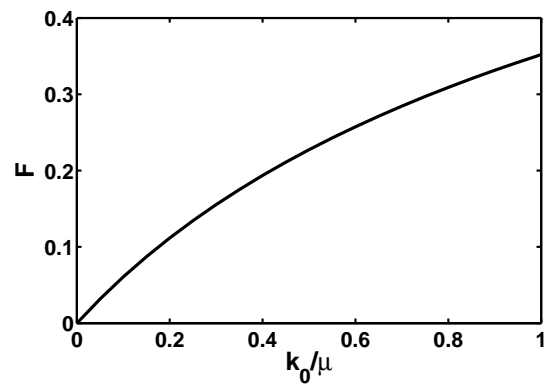}
  \caption{The function $F$ versus $k_0$ for a
  degenerate relativistic electron gas.
  \label{F}
  }
\end{figure}

Basing on Eqs.~\eqref{Fdeg} and~\eqref{nachdeg}, in Fig.~\ref{F} we plot the function
$F$ for a relativistic degenerate electron gas. One can see that $F\to0$ and thus $\Pi_{2}\to0$
for small $k_{0}$ in agreement with Eqs.~\eqref{eq:P21} and~\eqref{eq:J02p}.

%

Note that the nonzero $\Pi_{2}(0)=\Pi_{2}(k_{0}=0)$ can potentially
generate the instability of a large scale seed $B$-field resulting
in its exponential growth (see, e.g., Ref.~\cite{DvoSem14}). Longitudinal plasmons, contributing to the $B$-field growth, can be created in this case.
However, as results from Fig.~\ref{F}, in our situation $\Pi_{2}(0) = 0$ and $\tfrac{\mathrm{d}\Pi_{2}}{\mathrm{d}k_0}(k_{0}=0) \neq 0$. To study the $B$-field evolution in this case we should analyze the modified Maxwell equations
\begin{equation}\label{Maxeq}
  \mathrm{i}
  \left(
    \mathbf{q} \times \mathbf{B}
  \right) =
  - \mathrm{i} \omega \mathbf{E} + \sigma\mathbf{E} + \mathbf{j}_5,
  \quad
  \mathrm{i}
  \left(
    \mathbf{q} \times \mathbf{E}
  \right) =
  \mathrm{i} \omega \mathbf{B},
  \quad
  \left(
    \mathbf{q} \cdot \mathbf{B}
  \right) = 0,
\end{equation}
where $\sigma$ is the plasma conductivity, $\mathbf{E}$ is the electric field, and
\begin{equation}\label{j5}
  \mathbf{j}_5 = \Pi_2 (\omega) \mathbf{B} = - \mathrm{i} \zeta \omega \mathbf{B}.
\end{equation}
Here $\zeta$ is the constant parameter, which can be obtained using the definition of $F$ as
\begin{equation}\label{zeta}
  \zeta =
  \mathrm{i}\frac{\alpha_{\mathrm{em}}}{\sqrt{2}\pi}
  \left(
    1-4\sin^{2}\theta_{W}
  \right)
  \frac{G_{\mathrm{F}} (n_{e} - n_{\bar{e}})}{\mu}
  \left.
    \frac{\mathrm{d}F}{\mathrm{d}x}
  \right|_{x=0}.
\end{equation}
In Eqs.~\eqref{Maxeq} and~\eqref{j5} we use the Fourier representation of the electromagnetic field $\sim e^{- \mathrm{i} \omega t + \mathrm{i} \mathbf{qr}}$. It should be noted that $\zeta$ in Eq.~\eqref{zeta} is purely imaginary as predicted in Ref.~\cite{NiePal94} for a CP even Lagrangian, cf. Eq.~\eqref{eq:LFermi}.

Using the magnetohydrodynamic approximation,
\begin{equation}\label{MHDapp}
  \sigma \gg \omega,
  \quad
  |\mathbf{q}| |\mathbf{B}| \gg \omega |\mathbf{E}|,
\end{equation}
and assuming that
\begin{equation}\label{newapp}
  |\mathbf{B}| \gg -\mathrm{i} \zeta |\mathbf{E}|,
\end{equation}
on the basis of Eqs.~\eqref{Maxeq} and~\eqref{j5}, we get the equation for the $B$-field evolution,
\begin{equation}\label{neweveq}
  \mathbf{q}^2 \mathbf{B} =
  \mathrm{i} \sigma \omega\mathbf{B} - \zeta^2 \omega^2 \mathbf{B}.
\end{equation}
Using Eq.~\eqref{neweveq}, we obtain the dispersion relation,
\begin{equation}\label{omegaexp}
  \omega =
  \frac{\mathrm{i}}{2\zeta^{2}}
  \left(
    \sigma\mp\sqrt{\sigma^{2} + 4\zeta^{2}\mathbf{q}^{2}}
  \right).
\end{equation}
One can see that, at any values of $\zeta$, $\sigma$, and $|\mathbf{q}|$, $\mathrm{Im}(\omega) \leq 0$ in Eq.~\eqref{omegaexp}. Thus there is no $B$-field growth in the system in question. A seed magnetic field is attenuated instead.

Let us discuss the approximations made in deriving of Eq.~\eqref{neweveq}. First we note that $- \mathrm{i} \zeta \ll 1$ in the case of an electron plasma component in a magnetar. In this situation $n_{\bar{e}} = 0$. On the basis of Fig.~\ref{F} as well as Eqs.~\eqref{Fdeg} and~\eqref{nachdeg}, we get that $\left. \tfrac{\mathrm{d}F}{\mathrm{d}x} \right|_{x=0} \approx 0.5$. The chemical potential of a relativistic degenerate electron gas is $\mu = (3 \pi n_e)^{1/3}$. The electron density in a neutron star is maximal just after the core collapsing stage $n_e \sim 10^{37}\thinspace\text{cm}^{-3}$~\cite{DvoSem14}. At the subsequent stages of the magnetar evolution $n_e$ diminishes and reaches only a few percent of the neutron density. Assuming that $n_e = 10^{37}\thinspace\text{cm}^{-3}$ in Eq.~\eqref{zeta}, we get that $- \mathrm{i} \zeta \approx 4.5 \times 10^{-11} \ll 1$.

We shall study the situation when in Eq.~\eqref{neweveq} the $B$-field attenuation is small, i.e. $\sigma \ll - \mathrm{i} \zeta |\mathbf{q}|$. In this case the dispersion relation in Eq.~\eqref{omegaexp} reads, $|\mathbf{q}| = - \mathrm{i} \zeta \omega$. Therefore the condition in Eq.~\eqref{newapp} is automatically satisfied if the magnetohydrodynamic approximation in Eq.~\eqref{MHDapp} is valid. Indeed $|\mathbf{B}| \gg \tfrac{\omega}{|\mathbf{q}|} |\mathbf{E}|  = \tfrac{\mathrm{i}}{\zeta} |\mathbf{E}| \gg - \mathrm{i} \zeta |\mathbf{E}|$, since $- \mathrm{i} \zeta \ll 1$.

If we discuss a situation opposite to that in Eq.~\eqref{MHDapp}, i.e. assume that $\omega \gg \sigma$, one can obtain from Eqs.~\eqref{Maxeq} and~\eqref{j5} that $\mathbf{B}$ as well as $\omega$ and $\mathbf{q}$ obey the following equations:
\begin{equation}\label{bf}
  \mathbf{B} \pm \mathrm{i}
  \left(
    \mathbf{e}_q \times \mathbf{B}
  \right) = 0,
  \quad
  \omega^2 = |\mathbf{q}|^2
  \left(
    \frac{1}{\chi\epsilon} \pm
    \mathrm{i}
    \frac{\zeta\omega}{\epsilon|\mathbf{q}|}
  \right),
\end{equation}
where $\mathbf{e}_q = \mathbf{q}/|\mathbf{q}|$ is the unit vector. In Eq.~\eqref{bf} we also restore the nonzero permittivity $\epsilon$ and permeability $\chi$. Eq.~\eqref{bf} describes the birefringence of electromagnetic waves~\cite{NiePal94}. However, the amplitude of the $B$-field is constant, i.e. again no instability occurs.

Now let us make some general comments on the method adopted for the calculation of $\Pi_2$. The electromagnetic vertex $\gamma^\mu$ can get radiative corrections in the presence of $e^- e^+$ plasma with finite temperature and chemical potential. These corrections were studied in frames of QED in Ref.~\cite{DonHolRob85}.
The correction to $\gamma^\mu$ is additive and proportional to $\alpha_\mathrm{em} \approx 1/137$. Taking into account that $\Pi_2$ in Eq.~\eqref{eq:P21} is also proportional to $\alpha_\mathrm{em}$, we get that the corresponding correction to $\Pi_2$ is $\sim 10^2$ times smaller than the leading term in Eq.~\eqref{eq:P21}.

In the imaginary time perturbation theory used in our work the electron propagator $S_\mathrm{F}(k)$ is unchanged. Nevertheless, since we account for the dispersion relation of a plasmon $k^2 = k^2(T,\mu)$ in matter, we have to take into account the radiative correction to the electron mass $m \to m_\mathrm{eff}(T,\mu)$. Firstly, as shown in Refs.~\cite{DvoSem14,Bra92}, $m_\mathrm{eff}^2$ and $k^2$ are of the same order of magnitude. Therefore one should take them into account simultaneously in $\mathcal{E}_\mathbf{p}$ in Eqs.~\eqref{eq:P21} and~\eqref{eq:J02p}. Secondly, to avoid the plasmon decay into $e^- e^+$ pairs~\cite{Bra92}, we should guarantee that $k^2 < 4 m_\mathrm{eff}^2$. One can check that it is the case for a degenerate electron gas in a magnetar.

In conclusion we notice that we have explicitly demonstrated that
a medium consisting of electrons and positrons self-interacting within
the Fermi model, cf. Eq.~(\ref{eq:LFermi}), does not reveal an instability of a $B$-field leading to its growth. We have studied a particular case of a relativistic degenerate electrons inside a magnetar and showed that, using this mechanism, one cannot explain the amplification of a $B$-field of a protostar to the values observed in a magnetar contrary to the claim of Ref.~\cite{BoyRucSha12}.

We should note that the instability of the $B$-field does not exist in a hot relativistic $e^- e^+$ plasma, with $T\gg(m,\mu)$, either. To analyze this case one should take into account that $k^{2} \approx \tfrac{4\pi}{9}\alpha_{\mathrm{em}} T^{2}$, $k_{0}\approx0.1T\ll T$, and $m^2 \to m_{\mathrm{eff}}^{2}=\tfrac{e^{2}}{8} T^2$ (see Refs.~\cite{DvoSem14,Bra92}) in Eqs.~\eqref{eq:P21} and~\eqref{eq:J02p}. It
means that one cannot use this mechanism for the description of the $B$-field
amplification in the early universe at $m\ll T(\ll m_{\mu})$, where
$m_{\mu}$ is the muon mass.

\begin{acknowledgments}
I am thankful to V.~B.~Semikoz for helpful comments, to FAPESP
(Brazil) for a grant, and to Y.~S.~Kivshar for the hospitality at ANU where a part of the work was made.
\end{acknowledgments}

\end{document}